\documentclass[conference]{IEEEtran}
\IEEEoverridecommandlockouts
\usepackage{cite}
\usepackage{amsmath,amssymb,amsfonts}
\usepackage{algorithmic}
\usepackage{graphicx}
\usepackage{textcomp}
\usepackage{orcidlink}
\usepackage{gensymb}
\usepackage{xcolor}

\def\BibTeX{{\rm B\kern-.05em{\sc i\kern-.025em b}\kern-.08em
    T\kern-.1667em\lower.7ex\hbox{E}\kern-.125emX}}
\begin{document}

\title{Machine Learning Guided Cooling System Optimization for Data Center\\
\thanks{Won Prof. Avram Bar-Cohen Best Paper Award in the Data Centers Thermal Management track at 2026 IEEE ITherm.}
}

\author{
    \IEEEauthorblockN{Shrenik Jadhav}
    \IEEEauthorblockA{\textit{Department of Computer and Information Science}\\ 
    \textit{University of Michigan-Dearborn}\\
    Dearborn, USA \\
    \orcidlink{0009-0003-6906-7465}\url{https://orcid.org/0009-0003-6906-7465}}
    \and
    \IEEEauthorblockN{Zheng Liu}
    \IEEEauthorblockA{\textit{Department of Industrial and Manufacturing Systems Engineering}\\ 
    \textit{University of Michigan-Dearborn}\\
    Dearborn, USA \\
    \orcidlink{0000-0003-4869-8893}\url{https://orcid.org/0000-0003-4869-8893}}
}

\maketitle

\begin{abstract}
Effective data center cooling is crucial for reliable operation; however, cooling systems often exhibit inefficiencies that result in excessive energy consumption. This paper presents a three-stage, physics-guided machine learning framework for identifying and reducing cooling energy waste in high-performance computing facilities. Using one year of 10-minute resolution operational data from the Frontier exascale supercomputer, we first train a monotonicity-constrained gradient boosting surrogate that predicts facility accessory power from coolant flow rates, temperatures, and server power. The surrogate achieves a mean absolute error of 0.026 MW and predicts power usage effectiveness within $\pm$0.01 of measured values for 98.7\% of test samples. In the second stage, the surrogate serves as a physics-consistent baseline to quantify excess cooling energy, revealing approximately 85 MWh of annual inefficiency concentrated in specific months, hours, and operating regimes. The third stage evaluates guardrail-constrained counterfactual adjustments to supply temperature and subloop flows, demonstrating that up to 96\% of identified excess can be recovered through small, safe setpoint changes while respecting thermal limits and operational constraints. The framework yields interpretable recommendations, supports counterfactual analyses such as flow reduction during low-load periods and redistribution of thermal duty across cooling loops, and provides a practical pathway toward quantifiable reductions in accessory power. The developed framework is readily compatible with model predictive control and provides a template that, with site-specific recalibration, could be adapted to other liquid-cooled data centers with different configurations and cooling requirements.

\end{abstract}


\section{Introduction}
Modern high-performance computing (HPC) systems are now among the largest single loads in data centers. A single top-tier supercomputer can draw tens of megawatts of electrical power, and almost all that power eventually leaves the facility as low-grade heat that must be removed by the cooling plant \cite{iea2022datacenters, ebrahimi2014datacenter}. Even when an HPC facility reports an excellent Power Usage Effectiveness (PUE) of approximately 1.05, the absolute energy consumption of pumps, fans, and heat-rejection equipment remains extremely high every year \cite{greengrid2012pue}. Small percentage improvements in cooling efficiency, therefore, translate into meaningful operating cost savings and non-trivial reductions in emissions \cite{iea2022datacenters, gao2014machine}.

The Frontier supercomputer demonstrates a system that operates at high efficiency while retaining significant potential for further optimization. Frontier typically operates between 8 and 29 MW of IT (information technology) load, with an annual average of approximately 12 MW, and an observed PUE around 1.05 \cite{frontier2022factsheet}. In other words, the facility is already running near state-of-the-art efficiency, but that still corresponds to hundreds of kilowatts of continuous accessory power for cooling and infrastructure. Our exploratory data analysis confirms that cooling overhead becomes particularly visible during low-load periods, when fixed pump and fan power is amortized over relatively little compute work.

At the same time, data centers are under increasing pressure to support 
institutional sustainability goals and to unlock additional value from 
waste heat, for example, through district heating or heat pump 
integration \cite{yuan2023waste, irena2020waste}. Many of the remaining 
opportunities are not large design changes, but micro-optimizations: 
slightly raising supply temperatures, slightly reducing coolant flows, 
or rebalancing subloops while still respecting all thermal and 
reliability constraints \cite{ashrae2015thermal, doe2024bestpractices}. 
These are exactly the kinds of adjustments that operators are often 
reluctant to make manually, because the safe operating envelope is not 
obvious and the benefits can be smaller than day-to-day noise 
\cite{gao2014machine, dechiara2020datamining}. This motivates an approach that can mine historical data for such micro-inefficiencies, propose small, safe setpoint adjustments, and quantify their expected impact before any changes are made to the live system.

The study focuses on the cooling and power infrastructure serving the Frontier exascale system at Oak Ridge, which uses a warm water liquid cooling architecture with multiple secondary coolant loops that supply the compute cabinets and associated equipment \cite{frontier2022factsheet}. In this configuration, the IT racks reject heat into three main subloops, each with its own flow and return temperature measurements, while central pumps, heat exchangers, and towers remove that heat from the facility. The electrical power drawn by this infrastructure is reported as facility accessory power and, together with compute power, underpins both the site-level PUE and our definition of cooling overhead \cite{greengrid2012pue, frontier2022factsheet}.

Our analysis uses a full year of operational data from 2023, provided in the Frontier HPC \& Facility Data workbook at 10-minute resolution \cite{sun2024energy, grant2024frontier}. After cleaning and time-ordering, the dataset contains 49,869 timestamps spanning 2023-01-01 to 2023-12-31, slightly fewer than the 52,560 intervals in a complete year, reflecting scheduled downtime and short telemetry gaps. Each record includes synchronized measurements of per-loop coolant return temperatures and flow rates, overall supply temperature and flow, per-loop and total waste heat, compute power, facility accessory power, total power, and PUE, along with derived calendar fields (hour, month, weekday). Data quality of the dataset is high, with all power and flow channels are complete, and only a small fraction of overall return temperature readings are missing.
Table~\ref{tab:dataset-overview} summarizes the key variables and their characteristics in the Frontier dataset.

\begin{table}[h]
\centering
\caption{Overview of the Frontier HPC \& Facility Dataset.}
\label{tab:dataset-overview}

\begin{tabular}{lll}
\hline
\textbf{Variable} & \textbf{Description} & \textbf{Typical Range} \\
\hline
$P_{\text{IT}}$ & IT/compute power & 8--29 MW \\
$P_{\text{acc}}$ & Facility accessory power & 0.5--1.1 MW \\
$T_{\text{sup}}$ & Coolant supply temperature & 18--25$^{\circ}$C \\
$T_{r,i}$ & Subloop return temperatures & 25--40$^{\circ}$C \\
$Q_i$ & Subloop coolant flow rates & Variable \\
$Q_{\text{heat}}$ & Total waste heat & 5--25 MW \\
PUE & Power usage effectiveness & 1.03--1.10 \\
\hline
\multicolumn{3}{l}{Resolution: 10 minutes; Period: 2023-01-01 to 2023-12-31} \\
\multicolumn{3}{l}{Total records: 49,869 (after cleaning)} \\
\hline
\end{tabular}

\end{table}
A detailed exploratory data analysis on this dataset highlights several characteristic behaviors of the Frontier installation. Compute power and overall waste heat are tightly coupled, with average waste heat around 9 MW and rare peaks above 25 MW. PUE remains close to 1.05 for most of the year but rises noticeably at very low IT load, reflecting fixed cooling overheads. The three coolant subloops are not perfectly balanced: one loop carries the majority of the thermal load, while the others show periods of lower temperature and reduced flow. Finally, the distribution of waste heat is dominated by the 5 to 15 MW range with only modest daily and weekly seasonality, indicating a relatively steady but unevenly distributed source of recoverable heat over the year.

Prior work on energy efficiency in HPC and data centers has followed three main tracks. The first focuses on workload and scheduling policies, aiming to reduce energy consumption while respecting performance or quality-of-service constraints. Thermal-aware and energy-aware schedulers for HPC clusters adjust job placement and dynamic voltage and frequency scaling (DVFS) settings in response to temperature and power models, often using cyber-physical formulations that couple workload, cooling, and thermal dynamics \cite{auweter2014case, bhalachandra2015dynamic, bhalachandra2017improving}. Survey work on energy-aware scheduling emphasizes that energy is now the dominant operating cost for many HPC installations and highlights the tight coupling between temperature, cooling effort, and scheduling decisions \cite{lazic2018datacenter}. Recent spatio-temporal scheduling algorithms, which explicitly model thermal diffusion across the physical layout of data centers, leverage this structure to steer jobs away from hot spots and reduce cooling effort without violating thermal limits \cite{bhalachandra2017improving}.

The second major line of work focuses on cooling and plant aspects, employing advanced control to reduce the power of chiller, pump, and fan systems. Model-predictive control (MPC) has been successfully applied to data center chiller plants, using physics-based models and forecasts of load and weather to optimize water temperatures, flow rates, and equipment staging under operational constraints \cite{wang2022physics}. More recently, deep reinforcement learning (DRL) has been used to learn cooling policies directly from data, including algorithms that co-optimize IT and cooling systems \cite{moriyama2018reinforcement}, transform cooling optimization into a DRL control problem \cite{wang2025tropical}, jointly optimize job scheduling and cooling \cite{meta2024simulator}, and operate free-cooled data centers in tropical climates \cite{beloglazov2011energy}. These approaches demonstrate substantial energy savings, but they typically rely on complex black-box policies and are often validated on air-cooled facilities rather than warm water liquid-cooled supercomputers.

A third, emerging strand combines safety-aware learning with physics-guidance, recognizing that purely black-box controllers are difficult to deploy in critical infrastructure. Recent work on safe RL for building and plant control introduces explicit constraints, action-space pruning, and physics-inspired monotonicity checks to keep learned policies within safe operating envelopes \cite{wang2022physics, wang2025tropical}. Monotonicity constraints are particularly valuable because they encode domain knowledge that certain relationships should only increase or decrease: for example, higher coolant flow or greater heat load should never reduce cooling power, and enforcing this prevents the model from learning spurious correlations that could lead to unsafe recommendations. However, most of these studies either operate at the building-HVAC level or treat the data center as a single thermal zone; they rarely expose the fine-grained liquid-cooling loop behavior or accessory-power breakdown needed to reason about micro-optimizations such as small supply-temperature shifts or loop-specific flow trims. To our knowledge, the publicly released Frontier HPC \& facility dataset~\cite{sun2024energy, grant2024frontier} has not yet been studied based on physics-guided, guardrail-constrained micro-tuning of liquid-cooling operation at 10-minute resolution.

Against this backdrop, the specific gap we address is the lack of a transparent, physics-respecting framework that turns a year of high-resolution Frontier telemetry into safe, auditable micro-actions on the cooling plant. Concretely, we ask whether it is possible to accurately surrogate the accessory power as a function of IT load, temperatures, and flows; quantify when and where the excess cooling energy was used over the year; and construct a set of counterfactual, safety-screened tweaks to supply temperature and subloop flows that would have reduced that excess, while remaining close to historically observed operating conditions and producing reviewer-friendly logs of every proposed change. Filling this gap requires combining physics-guided modeling, conservative counterfactual evaluation, and explicit safety and plausibility checks, rather than relying on an opaque optimizer that operators would be reluctant to trust in an exascale production environment.

\section{Methodology}
This study implements a physics-guided, data-driven pipeline on top of 
the publicly released Frontier HPC \& facility dataset. 
Using measurements of IT power, coolant temperatures, coolant flow rate, 
waste heat, accessory power, and PUE at 10-minute resolution, we construct 
a three-stage workflow. In Step~1, we train a monotone, regime-aware surrogate 
model to predict accessory power from physics-inspired features. In Step~2, 
this surrogate defines an expected baseline; by comparing actual versus 
expected accessory power at each interval, we quantify excess cooling energy 
(MWh) and the associated cost. In Step~3, a counterfactual evaluation layer 
applies small, constrained changes to supply temperature and loop flows at 
each interval and estimates the associated energy and cost savings.

\begin{figure}[h]
    \centering
    \includegraphics[width=\columnwidth]{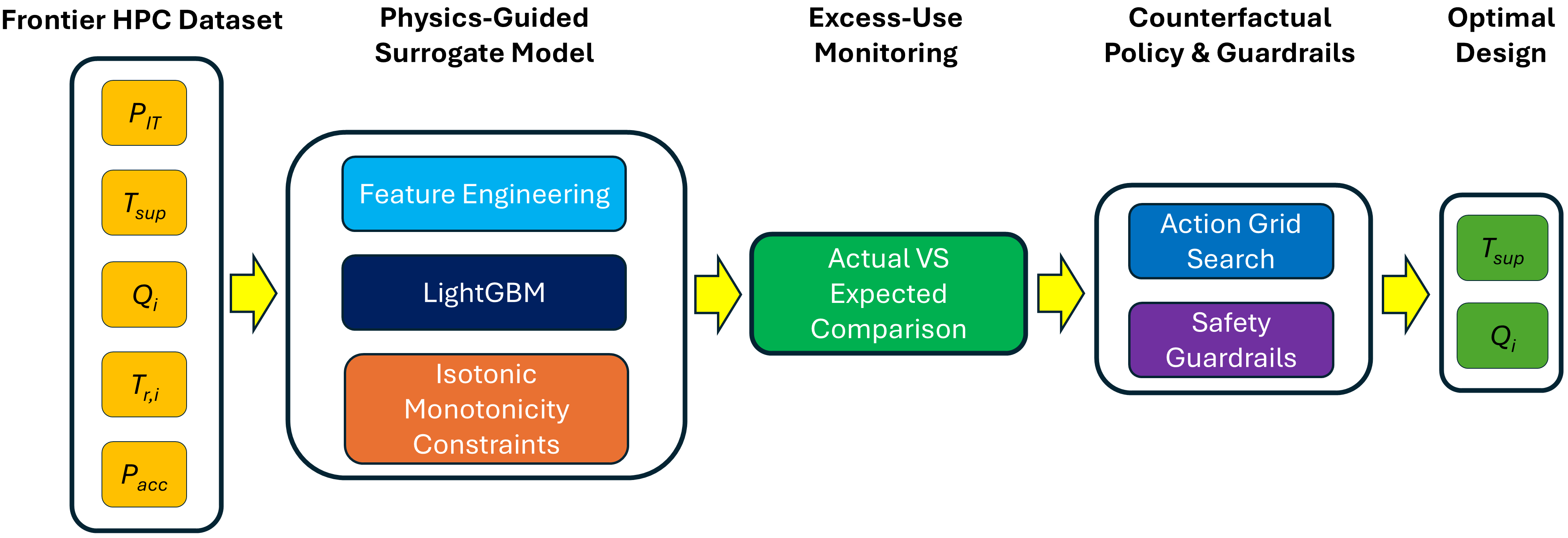}
    \caption{Flowchart of the physics-guided framework.}
    \label{fig:workflow}
\end{figure}

To make the results deployable, the three technical stages are wrapped in a reviewer-facing diagnostics layer. This layer enforces safety guardrails as discussed in Section \ref{sec:step3}, checks that counterfactual states remain inside the training distribution, applies materiality thresholds and simple ramp-rate/hysteresis filters, and produces an explicit action log for each 10-minute step. Figure~\ref{fig:workflow} summarizes the end-to-end framework from raw time series to surrogate modeling, excess-use characterization, counterfactual savings estimation, and reviewer-ready recommendations.

\subsection{Step 1. Physics-guided surrogate for accessory power}
The first stage constructs a supervised surrogate model for facility accessory power ($P_{\text{acc}}$) at 10-minute resolution. The model takes the IT load, liquid-cooling conditions, and simple temporal/context features as inputs and outputs a prediction of the accessory power consistent with the observed physics of the Frontier plant. This surrogate is then reused in Steps~2 and~3 as the digital twin baseline for cooling energy.

Feature design is directly informed by the coolant loop physics and by the EDA. Starting from the raw measurements of IT power ($P_{\text{IT}}$), overall supply temperature ($T_{\text{sup}}$), subloop return temperatures ($T_{r,1}, \ldots, T_{r,3}$), and subloop flows ($Q_1, \ldots, Q_3$), we compute per-loop temperature lifts as shown in Eq. \ref{eq:deltaT}.
\begin{equation}
    \Delta T_i = T_{r,i} - T_{\text{sup}}
    \label{eq:deltaT}
\end{equation}

Per-loop heat ($Q_{i,\text{heat}}$) using a standard water heat-capacity model, and aggregate quantities such as total flow (Eq. \ref{eq:q_tot}) and total heat (Eq. \ref{eq:q_tot_heat}).
\begin{equation}
    Q_{\text{tot}} = Q_1 + Q_2 + Q_3
    \label{eq:q_tot}
\end{equation}

\begin{equation}
    Q_{\text{tot,heat}} = Q_{1,\text{heat}} + Q_{2,\text{heat}} + Q_{3,\text{heat}}
    \label{eq:q_tot_heat}
\end{equation}

An imbalance index $I$ capturing how much of the heat is carried by the dominant loop. We add calendar features (hour, month) and low-load flags, as well as short-horizon history features (lags and rolling means of $P_{\text{IT}}$, $Q_{\text{tot}}$, and $T_{\text{sup}}$) to reflect the inertia of the cooling system. Finally, we fit a 3-cluster K-Means model on $(Q_{\text{tot}}, T_{\text{sup}})$ to obtain an operating regime label (0/1/2) per interval, which is passed as an additional categorical-like input to the surrogate.

We selected Light Gradient Boosting Machine (LightGBM) as the surrogate model for several reasons. First, gradient boosting methods handle mixed feature types (continuous, categorical, temporal) naturally without extensive preprocessing. Second, LightGBM natively supports monotonicity constraints on selected features, which is essential for encoding physics-based priors. Third, compared to deep learning approaches, LightGBM provides faster training, lower computational overhead, and more interpretable feature importance rankings, all of which are valuable for operational deployment in critical infrastructure. Finally, tree-based ensembles have demonstrated strong performance on tabular regression tasks with moderate dataset sizes, making them well-suited to this 10-minute resolution telemetry problem.

We then train a Light Gradient Boosting Machine (LightGBM) with monotonicity constraints on key features that should only increase accessory power under normal physics (e.g., $P_{\text{IT}}$, $Q_{\text{tot}}$, $Q_{\text{tot,heat}}$, and related smoothed terms). This enforces that the learned mapping respects basic trends such as more heat and more flow cannot reduce cooling power, which is important for later guardrail design. The model is trained on an 80/20 random split of the 2023 time series, with early stopping on a held-out slice of the training set. To correct any residual calibration bias, we apply one-dimensional isotonic regression to the raw LightGBM outputs, yielding a final calibrated prediction $\hat{P}_{\text{acc}}$. From this, an implied PUE can be computed with values explicitly clipped to be $\geq 1.0$ in later steps as shown in Eq. \ref{eq:pue_hat}.
\begin{equation}
    \widehat{\text{PUE}} = \frac{P_{\text{IT}} + \hat{P}_{\text{acc}}}{P_{\text{IT}}}
    \label{eq:pue_hat}
\end{equation}

We evaluate the surrogate using standard regression metrics: Mean Absolute Error (MAE), Root Mean Square Error (RMSE), coefficient of determination ($R^2$), Symmetric Mean Absolute Percentage Error (SMAPE), Weighted Absolute Percentage Error (WAPE), and Root Mean Square Logarithmic Error (RMSLE) on both train and test splits, as well as PUE-space errors (MAE and RMSE of PUE, and the fraction of points where predicted PUE is within 0.01 of the measured value). The resulting test-set errors are small relative to typical accessory power and PUE variation, indicating that the surrogate is accurate enough to support the excess-use analysis (Step~2) and counterfactual policy evaluation (Step~3), while remaining simple, monotone, and interpretable.

\subsection{Step 2. Excess-use monitoring}
In the second stage, the surrogate from Step~1 is used as a physics-consistent baseline to identify intervals where the actual plant appears to use excess cooling power. For every 10-minute interval in 2023, we apply the trained, calibrated model from Step~1 to the realized operating conditions and obtain an expected accessory power $\hat{P}_{\text{acc}}$. Comparing this to the measured accessory power $P_{\text{acc,actual}}$ yields a residual that we interpret as potential excess use. To remain conservative, we define the excess cooling power at each time step such that intervals where the plant outperforms the model (negative residuals) are set to zero, rather than treated as negative excess or credited, as shown in Eq. \ref{eq:pexcess}.
\begin{equation}
    P_{\text{excess}}(t) = \max\bigl(P_{\text{acc,actual}}(t) - \hat{P}_{\text{acc}}(t),\, 0\bigr)
    \label{eq:pexcess}
\end{equation}

Because the data are sampled at a fixed 10-minute cadence, each interval corresponds to a duration that can be calculated as shown in Eq.~\ref{eq:deltat}.
\begin{equation}
    \Delta t = \frac{10}{60}\,\text{h} \approx 0.1667\,\text{h}
    \label{eq:deltat}
\end{equation}

We convert power to energy, as shown in Eq. \ref{eq:eexcess}.
\begin{equation}
    E_{\text{excess}}(t) = P_{\text{excess}}(t)\,\Delta t \; [\text{MWh}]
    \label{eq:eexcess}
\end{equation}

Given a specified electricity tariff, either flat or time-of-use, with price $c(t)$ in \$/kWh, we compute an interval-level cost of excess as shown in Eq \ref{eq:cexcess}.
\begin{equation}
    C_{\text{excess}}(t) = E_{\text{excess}}(t) \times 1000 \times c(t)
    \label{eq:cexcess}
\end{equation}

And optionally add a simple demand-charge term based on the monthly peak of $P_{\text{excess}}$ if a demand rate (in \$/MW per month, 
a standard utility tariff structure) is provided. This yields a time series of excess power, energy, and cost that is consistent with the Step 1 physics-guided baseline and the chosen tariff structure.

To make this information operationally useful, we aggregate the interval-level excess into several views: (i) daily and monthly roll-ups of $E_{\text{excess}}$ and $C_{\text{excess}}$; (ii) an hour-by-month matrix showing average excess MWh per interval as a function of local time-of-day and season; and (iii) groupings by operating regime, using the regime labels from Step~1. Additional calendar features, such as day-of-week, are also retained to probe weekly patterns. These summaries allow us to identify when (which months, days, and hours) and under which regimes excess cooling energy is most pronounced, and they provide an upper bound on the savings that any subsequent counterfactual policy (Step~3) can credibly capture at each timestamp.

\subsection{Step 3. Counterfactual policy and reviewer diagnostics}\label{sec:step3}
In the third stage, the surrogate from Step~1 is used as a counterfactual oracle. The core idea is to ask: ``What if we had operated with slightly different setpoints at each historical timestamp?'' By replaying the year with small, hypothetical adjustments to supply temperature and flow rates, we can estimate the energy that could have been saved without actually modifying the live system. The guardrails ensure that these hypothetical adjustments remain physically plausible (e.g., heat must still be removed) and operationally safe (e.g., temperatures and flows stay within proven bounds). At each 10-minute interval, we estimat the accessory-power reduction that would have been achievable under a minimally perturbed cooling action, subject to the same safety constraints. To keep the policy interpretable and operationally realistic, we restrict attention to two familiar control knobs: the coolant supply temperature $T_{\text{sup}}$, and the per-loop coolant flows $Q_1, Q_2, Q_3$.

For each time step $t$, we construct a discrete action grid over small supply-temperature increases $\Delta T_{\text{sup}} \in \{0.0, 0.2, 0.4, \ldots, 1.5\}\,^{\circ}\text{C}$
subject to a hard cap $T_{\text{sup}}^{\max}$, and over multiplicative flow scales for the three subloops. The dominant loop (the loop carrying the largest fraction of baseline heat at that timestamp) is allowed slightly stronger reductions than the non-dominant loops, but no loop is ever allowed below $90\%$ of its baseline flow. For each candidate action $a$ in this grid, we form a counterfactual supply temperature where the supply temperature is capped at the maximum allowable value as shown in Eq. \ref{eq:Tsup_cf}.
\begin{equation}
    T_{\text{sup}}^{\text{cf}}(t,a) = \min\bigl(T_{\text{sup}}(t) + \Delta T_{\text{sup}}(a),\, T_{\text{sup}}^{\max}\bigr)
    \label{eq:Tsup_cf}
\end{equation}

We then scale the flows while holding fixed IT power $P_{\text{IT}}(t)$, return temperatures $T_{r,i}(t)$, calendar variables (hour, month), and all history features (lags and moving averages) for that 10-minute step as shown in Eq. \ref{eq:Qi_cf}.
\begin{equation}
    Q_i^{\text{cf}}(t,a) = s_i(a)\,Q_i(t)
    \label{eq:Qi_cf}
\end{equation}

From $\{T_{\text{sup}}^{\text{cf}}, Q_i^{\text{cf}}, T_{r,i}\}$ we recompute the physics-derived features: per-loop temperature lifts (Eq. \ref{eq:DeltaTi_cf}).
\begin{equation}
    \Delta T_i^{\text{cf}}(t,a) = \max\bigl(T_{r,i}(t) - T_{\text{sup}}^{\text{cf}}(t,a),\, \Delta T_{\min}\bigr)
    \label{eq:DeltaTi_cf}
\end{equation}

Also, we recompute per-loop and total heat $(Q_{i,\text{heat}}^{\text{cf}}, Q_{\text{tot,heat}}^{\text{cf}})$, total flow $Q_{\text{tot}}^{\text{cf}}$, nonlinear terms such as $\bigl(Q_{\text{tot}}^{\text{cf}}\bigr)^3$, and interactions such as $P_{\text{IT}}(t)\,T_{\text{sup}}^{\text{cf}}(t,a)$ and $Q_{\text{tot}}^{\text{cf}}(t,a)\,\overline{\Delta T}^{\text{cf}}(t,a)$. We then reevaluate the operating regime by applying the Step 1 K-means model to $\bigl(Q_{\text{tot}}^{\text{cf}}(t,a), T_{\text{sup}}^{\text{cf}}(t,a)\bigr)$, so that each counterfactual point is embedded in the same regime structure as the training data. This yields a counterfactual feature vector $x^{\text{cf}}(t,a)$, which is passed through the LightGBM booster and isotonic calibrator to obtain a predicted accessory power $P_{\text{acc}}^{\text{cf}}(t,a)$ and corresponding PUE as shown in Eq. \ref{eq:PUE_cf}.
\begin{equation}
    \text{PUE}^{\text{cf}}(t,a)
    = \frac{P_{\text{IT}}(t) + P_{\text{acc}}^{\text{cf}}(t,a)}{P_{\text{IT}}(t)}
    \label{eq:PUE_cf}
\end{equation}

Each candidate action is then filtered through explicit guardrails:
\begin{itemize}
    \item No unphysical efficiency: $\text{PUE}^{\text{cf}}(t,a) \ge 1$;
    \item Cooling preserved: $Q_{\text{tot,heat}}^{\text{cf}}(t,a) \ge \alpha\,Q_{\text{tot,heat}}(t)$ with $\alpha = 0.97$;
    \item Minimum thermal lift: $\Delta T_i^{\text{cf}}(t,a) \ge \Delta T_{\min}$ for all loops $i$;
    \item Minimum flow: $Q_i^{\text{cf}}(t,a) \ge 0.9\,Q_i(t)$ for all loops $i$;
    \item Supply-temperature cap: $T_{\text{sup}}^{\text{cf}}(t,a) \le T_{\text{sup}}^{\max}$.
\end{itemize}

Actions that pass all guardrails are considered feasible. For each feasible action, we define a conservative, single-step saving in MW as shown in Eq. \ref{eq:S_ta}.
\begin{equation}
    S(t,a) = \max\bigl(P_{\text{acc,actual}}(t) - P_{\text{acc}}^{\text{cf}}(t,a),\, 0\bigr)
    \label{eq:S_ta}
\end{equation}

At each time step, we select the action $a^{\star}(t)$ that maximizes $S(t,a)$, with tie-breaking in favor of smaller $\Delta T_{\text{sup}}$ and gentler flow reductions. The chosen saving is converted to energy (Eq. \ref{eq:E_save}) and cost (Eq. \ref{eq:C_save}) where $\Delta t$ is the 10-minute interval in hours and $c(t)$ is either a flat or time-of-use tariff. Aggregating $\{E_{\text{save}}(t), C_{\text{save}}(t)\}$ over time yields annual counterfactual savings, which we report by month, hour of day, operating regime, and dominant loop.
\begin{equation}
    E_{\text{save}}(t) = S\bigl(t,a^{\star}(t)\bigr)\,\Delta t
    \label{eq:E_save}
\end{equation}
\begin{equation}
    C_{\text{save}}(t) = E_{\text{save}}(t) \times 1000 \times c(t)
    \label{eq:C_save}
\end{equation}

On top of this counterfactual layer, a reviewer-oriented diagnostics module provides safety, plausibility, and materiality assessments in a form suitable for audit. First, for the selected actions it reconstructs the key counterfactual features $(T_{\text{sup}}^{\text{cf}}, Q_{\text{tot}}^{\text{cf}}, \overline{\Delta T}^{\text{cf}}, Q_{\text{tot,heat}}^{\text{cf}})$ and checks that they lie within the empirical 1st-99th percentile range of the Step 1 training distribution, yielding an in-distribution coverage percentage for each feature. Second, it summarizes any guardrail breaches (PUE, flows, temperature lift, temperature caps); in our experiments, the search is configured so that these counts are identically zero, i.e., no recommended action violates the encoded physics and safety constraints.

Third, the diagnostics introduce a materiality threshold for single-step  (Eq. \ref{eq:materiality}), so that candidate actions smaller than both an absolute floor $S_{\min}$ (MW) and half the surrogate’s test MAE are treated as noise rather than actionable savings. This avoids over-interpreting tiny differences within the model’s error band. To connect Step 3 savings back to the ``excess'' identified in Step~2, the module also computes a capped-capture metric (Eq. \ref{eq:E_save_cap}) and reports both raw and capped totals and capture rates (e.g., the fraction of Step 2 excess energy that is recoverable under this policy).
\begin{equation}
    S\bigl(t,a^{\star}(t)\bigr) \ge \max\bigl(S_{\min},\, 0.5 \times \text{MAE}_{\text{test}}\bigr)
    \label{eq:materiality}
\end{equation}

\begin{equation}
    E_{\text{save}}^{\text{cap}}(t)
    = \min\bigl(E_{\text{save}}(t),\, E_{\text{excess}}(t)\bigr)
    \label{eq:E_save_cap}
\end{equation}

Finally, simple operational realism checks are provided: action frequency, i.e., the fraction of intervals with $S\bigl(t,a^{\star}(t)\bigr) > 0$; ramp-rate indicators, based on step to-step changes in $T_{\text{sup}}$ and the chosen flow scales; and hysteresis indicators, enforcing a minimum number of idle steps between consecutive actions to avoid rapid toggling. Together, the counterfactual engine and diagnostics layer provide not only point estimates of potential savings, but also a structured, physics-aware justification for each recommended tweak and a set of aggregate metrics that reviewers can use to judge safety, plausibility, and operational relevance.

\section{Experimental Results \& Discussion}
\subsection{Surrogate performance}
\label{subsec:step1_surrogate}

The physics-guided LightGBM surrogate reproduces Frontier’s accessory power $P_{\mathrm{acc}}$ with errors on the order of only a few tens of kilowatts. On a random 80/20 split of the 2023 dataset, the test-set MAE is $0.0259~\mathrm{MW}$ and the RMSE is $0.0387~\mathrm{MW}$, with $R^{2}=0.79$ and $\mathrm{SMAPE}\approx 4.2\%$. The weighted absolute percentage error (WAPE) is $4.3\%$, and the log-scaled error (RMSLE) is $0.0233$. Training errors are slightly lower (MAE $0.0193~\mathrm{MW}$, RMSE $0.0293~\mathrm{MW}$, $R^{2}=0.88$), indicating a modest but expected generalization gap without evidence of severe overfitting. When mapped into PUE, the test-set error remains very small ($\mathrm{PUE\_MAE}=0.00225$, $\mathrm{PUE\_RMSE}=0.00406$), with $98.7\%$ of test points within $\pm 0.01$ of the measured PUE (Table~\ref{tab:surrogate-metrics}).
\begin{table}[h]
\centering
\caption{Performance of the physics-guided LightGBM surrogate for accessory power
$P_{\mathrm{acc}}$ on a random 80/20 train-test split.}
\label{tab:surrogate-metrics}
\begin{tabular}{lcc}
\hline
\textbf{Metric} & \textbf{Train} & \textbf{Test} \\
\hline
MAE [MW]              & 0.0193  & 0.0259 \\
RMSE [MW]             & 0.0293  & 0.0387 \\
$R^{2}$               & 0.8785  & 0.7909 \\
Adjusted $R^{2}$      & 0.8785  & 0.7909 \\
SMAPE [\%]            & 3.13    & 4.18   \\
WAPE                  & 0.0323  & 0.0432 \\
RMSLE                 & 0.0175  & 0.0233 \\
PUE MAE               & 0.00176 & 0.00225 \\
PUE RMSE              & 0.00430 & 0.00406 \\
PUE within $\pm 0.01$ & 0.9925  & 0.9871 \\
\hline
\end{tabular}
\end{table}
\subsection{Excess characterization}
\label{subsec:step2_excess}

For context, Frontier’s accessory power is typically in the range $0.6$–$0.8~\mathrm{MW}$, i.e., only a few percent of the total facility load. An MAE of $\sim 0.026~\mathrm{MW}$ therefore corresponds to errors on the order of a few tens of kilowatts and an average relative error of about $4\%$. The moderate test-set $R^{2}$ of $0.79$ mainly reflects the limited dynamic range and measurement noise of $P_{\mathrm{acc}}$, rather than large absolute errors: when the signal itself spans less than a megawatt, even small fluctuations in the target reduce the fraction of variance explained.  Importantly, the absolute error magnitude (MAE $\approx 0.026~\mathrm{MW}$) is substantially smaller than the savings claimed in Step~3, ensuring that the counterfactual conclusions are not dominated by model uncertainty. The percentage-based metrics (SMAPE, WAPE) and the very low PUE error provide a more operationally meaningful view and indicate that the surrogate is accurate enough for detecting deviations and evaluating sub-$0.1~\mathrm{MW}$ savings.

\begin{figure}[t]
  \centering
  \includegraphics[width=0.95\linewidth]{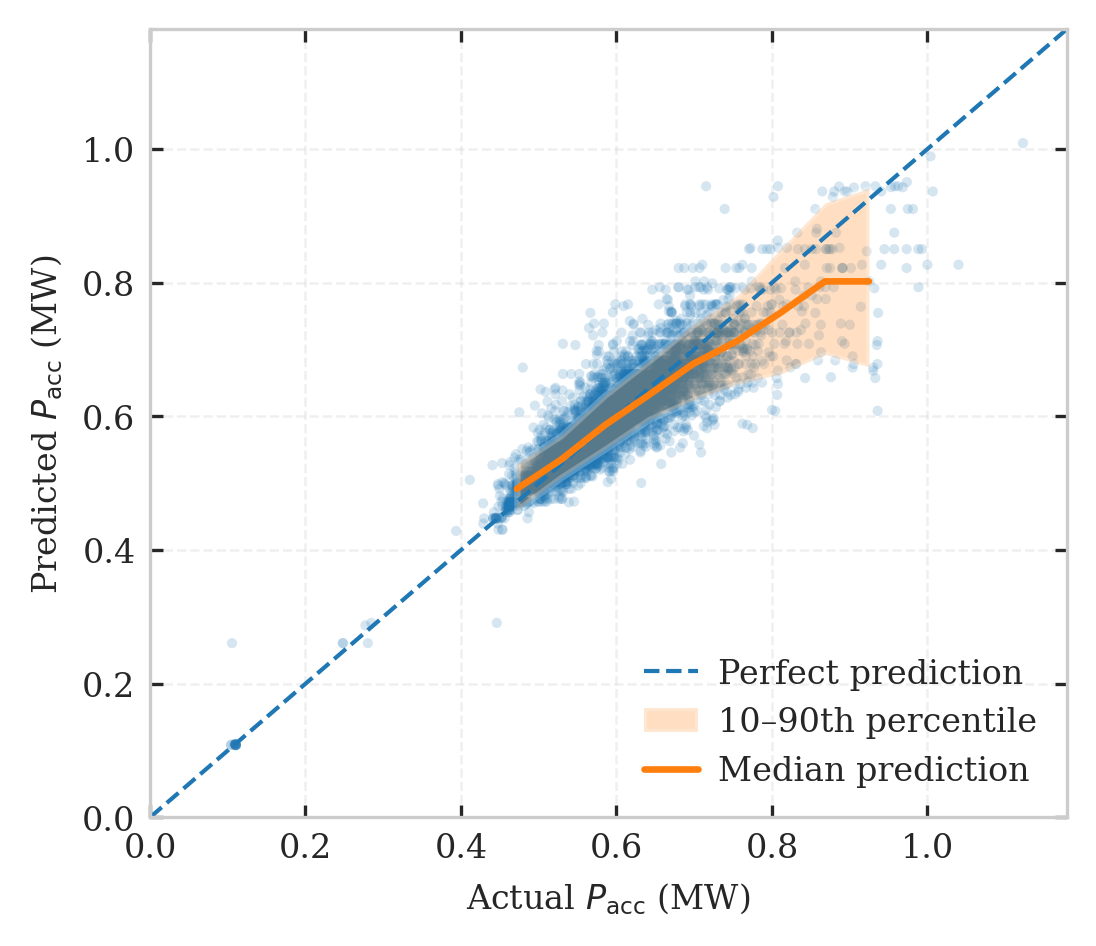}
  \caption{Calibration plot for the accessory-power surrogate on the 2023 test set. Each point is a 10-minute interval; the dashed line denotes perfect prediction, and the solid line with shaded band shows the binned median and 10th--90th percentile of predicted $P_{\mathrm{acc}}$ as a function of the actual value.}
  \label{fig:surrogate_parity}
\end{figure}

The calibration / parity plot (Fig.~\ref{fig:surrogate_parity}) shows that predictions lie close to the $45^{\circ}$ line across the full range of observed accessory powers; the median prediction curve closely follows the ideal line, and the 10th–90th percentile band remains narrow.

\begin{figure}[h]
  \centering
  \includegraphics[width=0.95\linewidth]{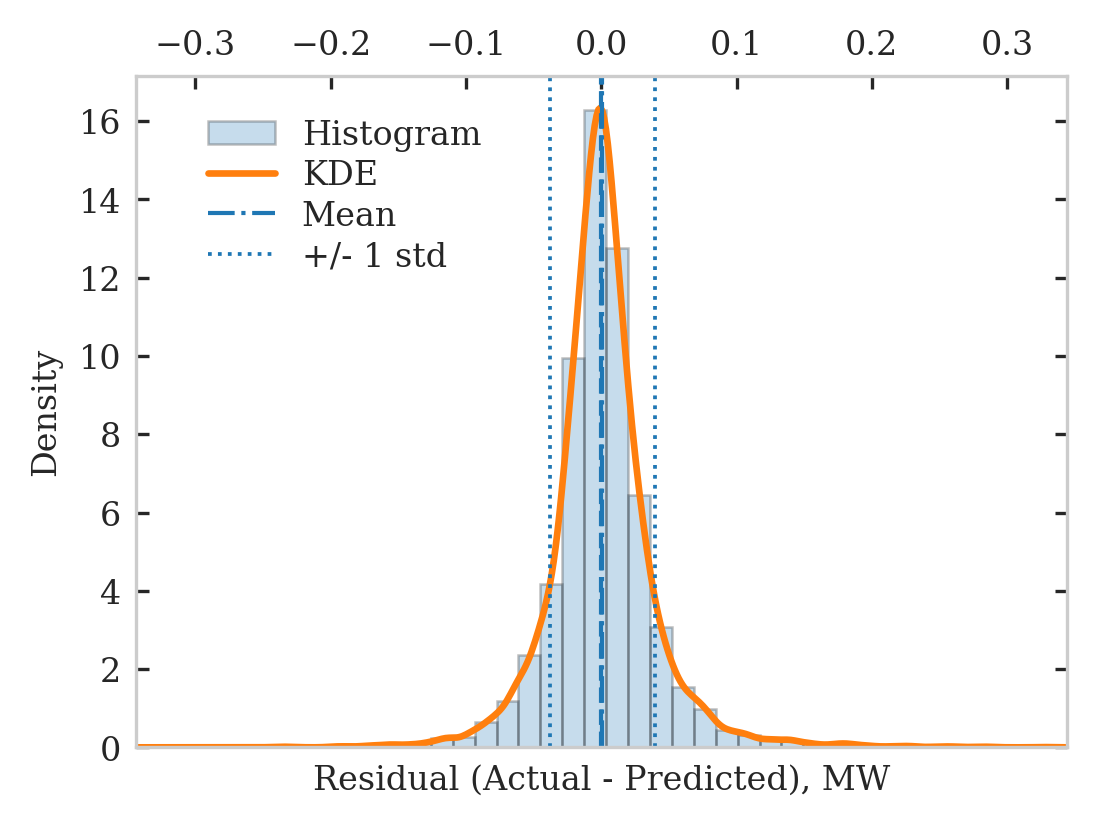}
  \caption{Residual distribution for the accessory-power surrogate on the test set. The histogram and KDE are approximately symmetric and centered near zero, with most residuals confined to a narrow band around the origin.}
  \label{fig:resid_dist}
\end{figure}

Residual histograms (Fig.~\ref{fig:resid_dist}) reveal an approximately symmetric residual distribution centered near zero (mean 0.0007~MW; median -0.0006~MW) with a standard deviation of 0.0387~MW and interquartile range of 0.0338~MW. Approximately 95\% of residuals lie within [-0.075, 0.084]~MW. Most errors are operationally small: 86.4\% of samples satisfy $|e|\leq 0.05~\mathrm{MW}$ (and 55.9\% satisfy $|e|\leq 0.02~\mathrm{MW}$).

\begin{figure}[h]
  \centering
  \includegraphics[width=0.95\linewidth]{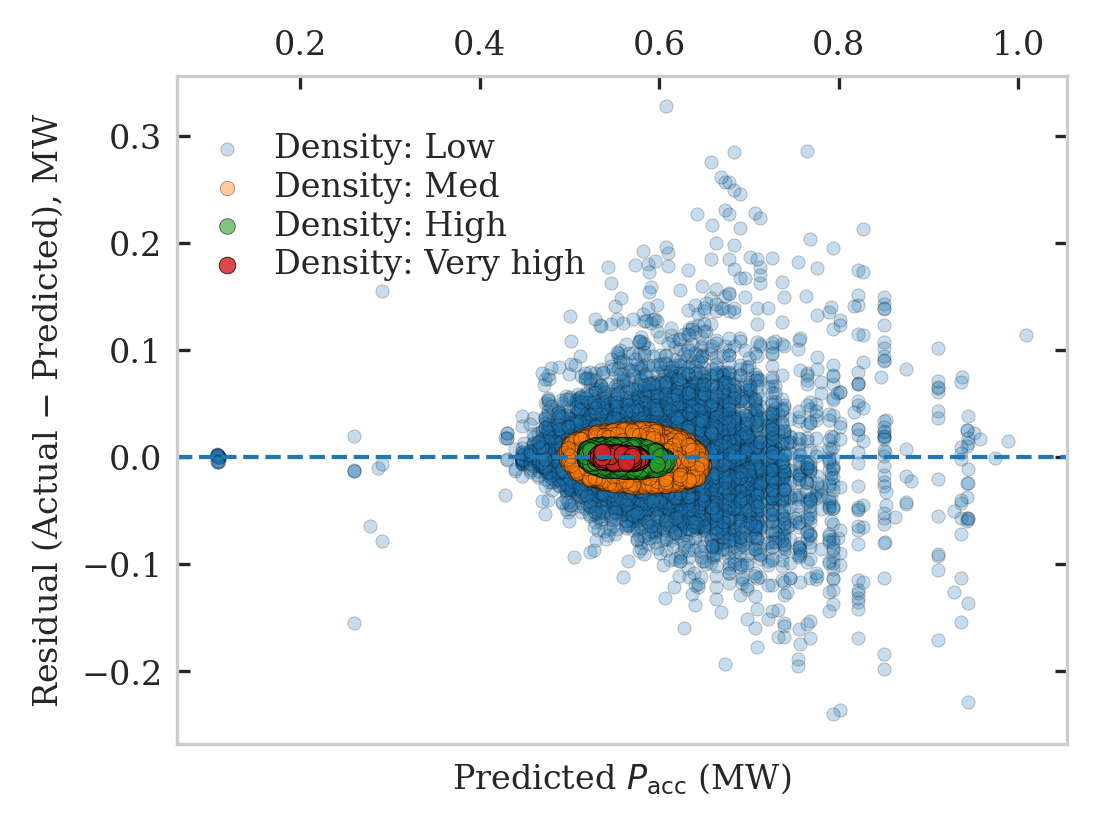}
  \caption{Residuals versus predicted accessory power on the test set. Residuals remain centered around zero across the prediction range and show no clear structure or heteroscedasticity.}
  \label{fig:resid_scatter}
\end{figure}

Residuals versus predicted values (Fig.~\ref{fig:resid_scatter}) do not exhibit obvious structure or heteroscedasticity, suggesting that the model does not systematically under- or over-predict in specific operating regions.

\begin{figure*}[h]
    \centering
    \includegraphics[width=0.75\textwidth]{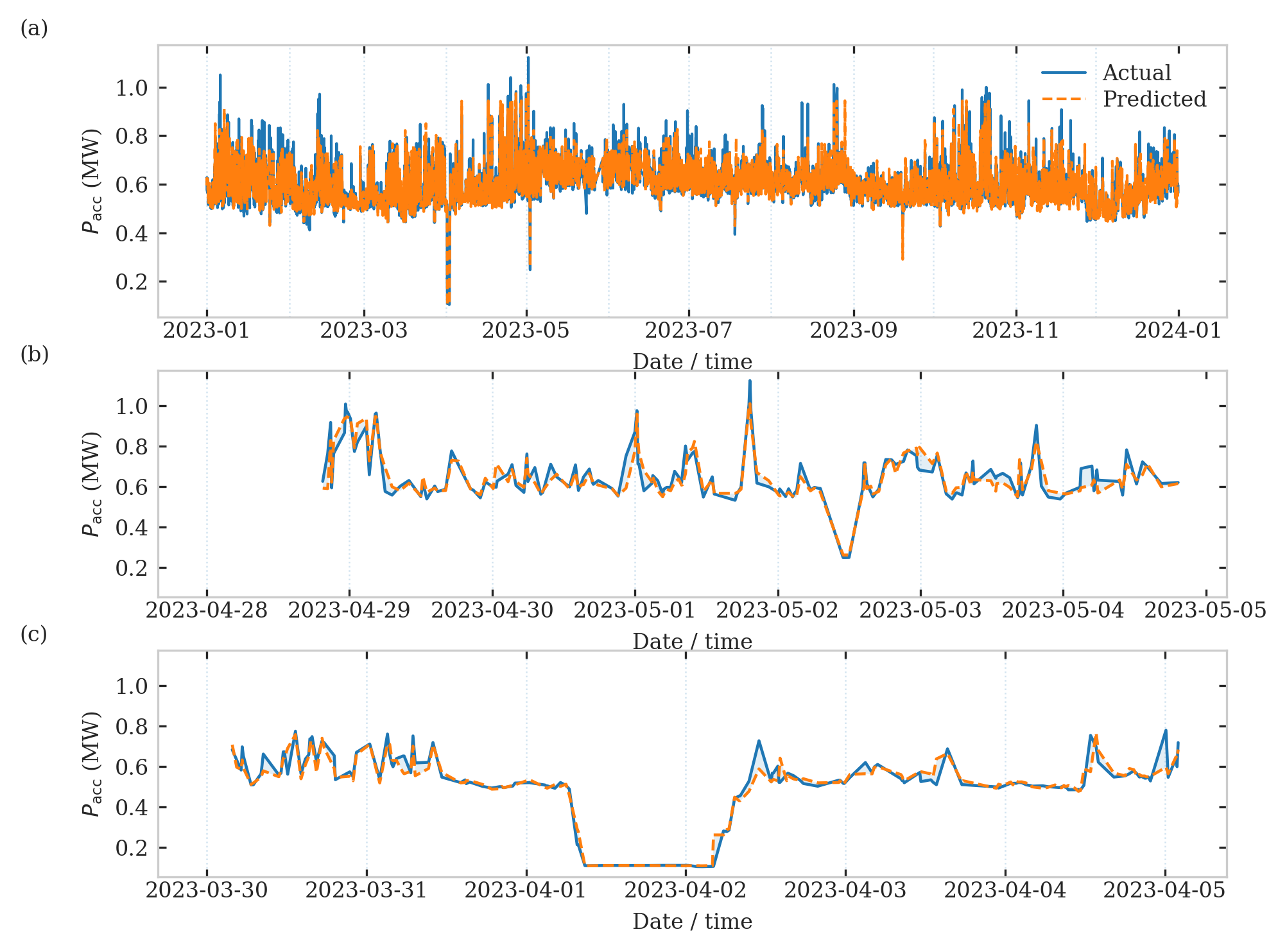}
    \caption{Time-series comparison of measured and predicted accessory power on the test set. (a) Full test period. (b) Zoom into a representative high-load week. (c) Zoom into a representative low-load / maintenance week.}
    \label{fig:timeseries_fit}
\end{figure*}

\begin{figure}[h]
  \centering
  \includegraphics[width=0.95\linewidth]{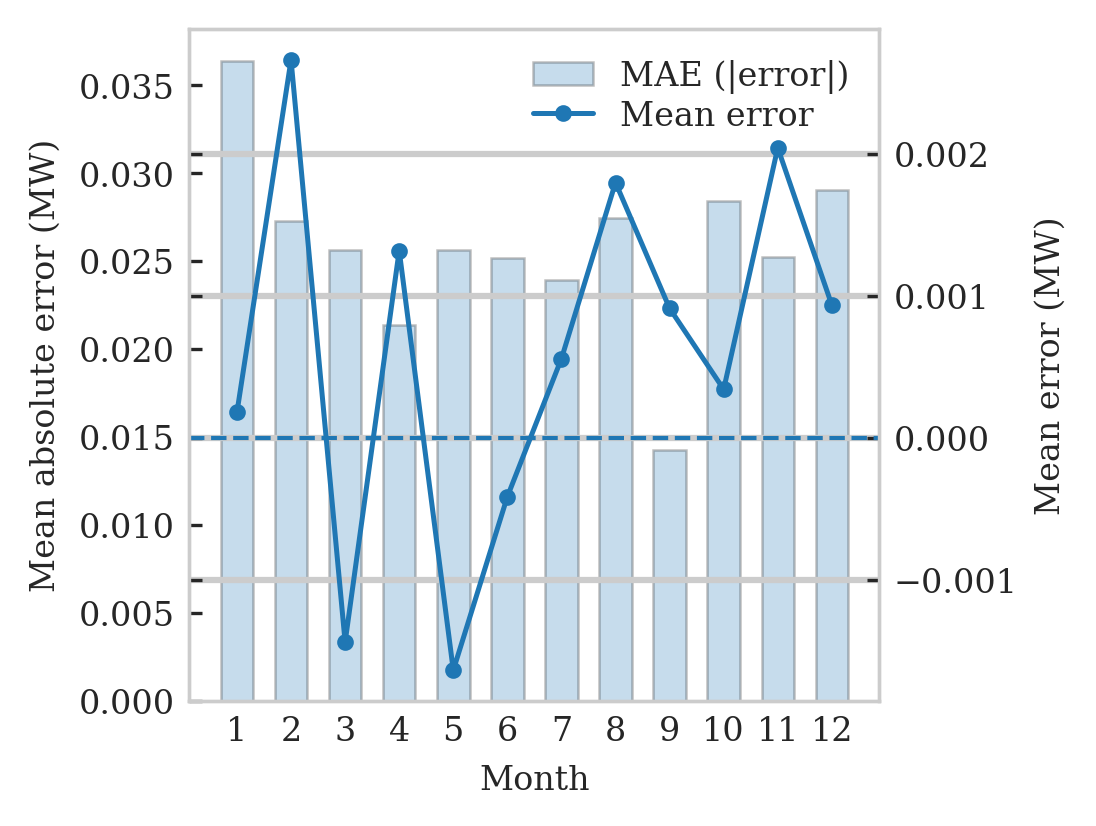}
  \caption{Monthly error profile of the accessory-power surrogate on the test set. Bars show mean absolute error (MAE) and the line shows signed mean error (bias), indicating stable error magnitude and negligible seasonal bias.}
  \label{fig:monthly_error}
\end{figure}

Time-series overlays for the test set (Fig.~\ref{fig:timeseries_fit}) confirm that the surrogate tracks 10-minute dynamics during both high-load periods and low-load / maintenance episodes without noticeable drift, while the monthly error profile (Fig.~\ref{fig:monthly_error}) shows MAE values in a tight band ($\sim 0.02$–$0.03~\mathrm{MW}$) and negligible seasonal bias.

Taken together, these results indicate that the Step 1 surrogate is well calibrated, nearly unbiased, and temporally stable at the tens-of-kilowatts level. This level of fidelity is sufficient to (i) define a robust ``expected'' baseline for accessory power in Step~2 and (ii) support the evaluation of the small, guardrail-constrained counterfactual adjustments explored in Step~3 without the conclusions being dominated by model error.

Using the Step 1 surrogate as a physics-consistent baseline, we quantify excess accessory power at 10-minute resolution over the 2023 Frontier dataset and then aggregate it to energy and cost. Over the full year, the implied excess cooling energy is $85.2~\mathrm{MWh}$, corresponding to an energy cost of approximately \$5,100 at \$60/MWh and a small additional demand charge, for a total annual excess cost of about \$5,100 (Table~\ref{tab:excess-summary}). Although this represents only a fraction of the overall cooling energy, it is a non-negligible resource that can be targeted by micro-optimizations.

\begin{table}[h]
\centering
\caption{Annual excess cooling energy and cost relative to the surrogate baseline.}
\label{tab:excess-summary}
\begin{tabular}{lc}
\hline
\textbf{Quantity} & \textbf{Value} \\
\hline
Total excess energy & $85.2~\mathrm{MWh}$ \\
Energy cost (@ \$60/MWh) & \$5,100\\

Total excess cost & $\approx\$5{,}100$ \\
\hline
\end{tabular}
\end{table}

\begin{figure*}[h]
  \centering
  \includegraphics[width=0.75\textwidth]{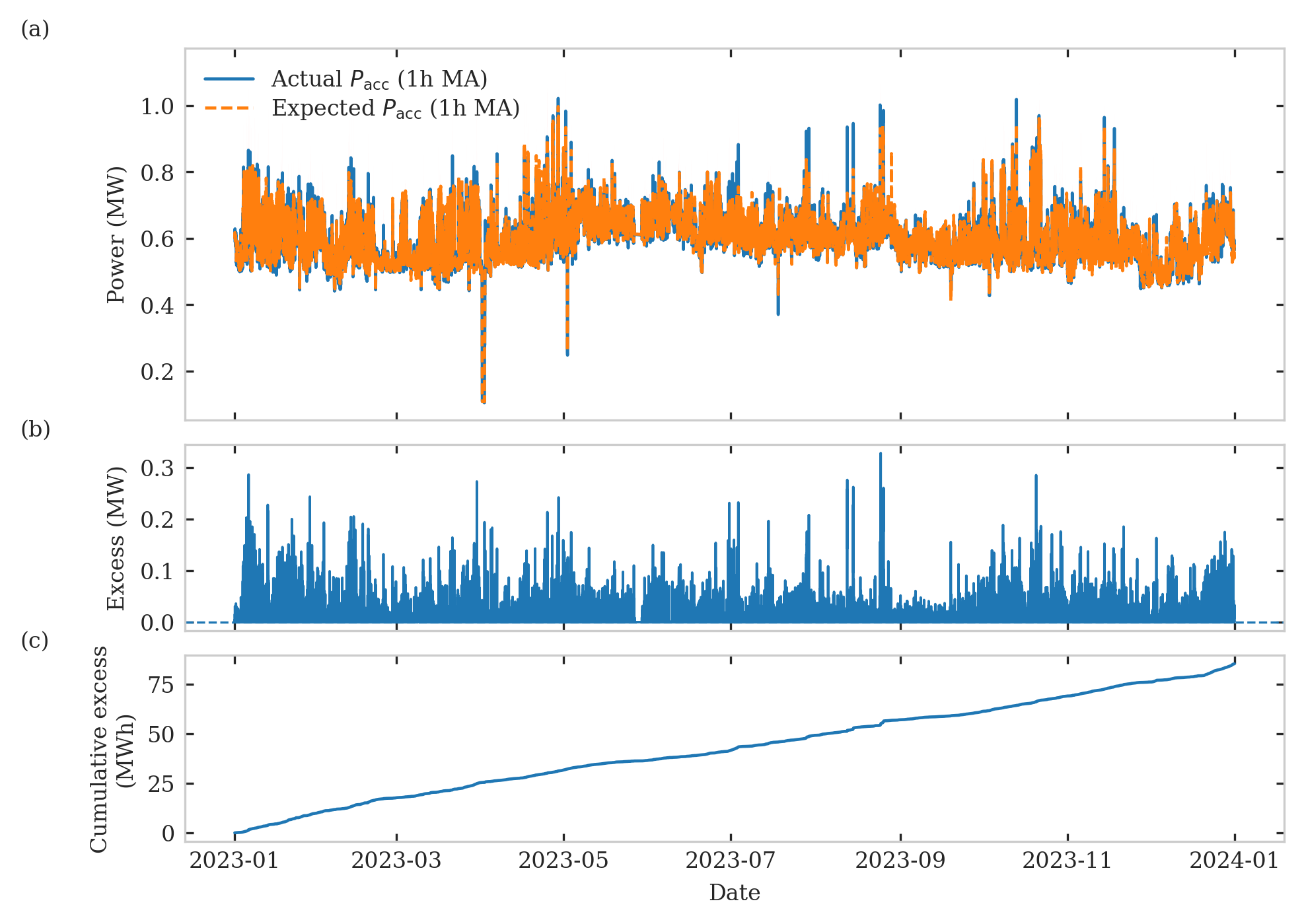}
  \caption{Accessory power relative to the surrogate baseline in 2023. (a) Actual and expected accessory power $P_{\mathrm{acc}}$ (smoothed). (b) Instantaneous excess cooling power. (c) Cumulative excess cooling energy.}
  \label{fig:excess_timeseries}
\end{figure*}

Figure~\ref{fig:excess_timeseries} illustrates how this excess arises over time. The upper panel compares actual and surrogate-expected accessory power $P_{\mathrm{acc}}$; intervals where the plant draws more power than expected. The middle panel shows the instantaneous excess cooling power, and the lower panel shows the cumulative excess energy, which grows approximately monotonically through the year, with visibly steeper segments during certain weeks in summer and winter.

\begin{figure}[h]
  \centering
  \includegraphics[width=0.9\linewidth]{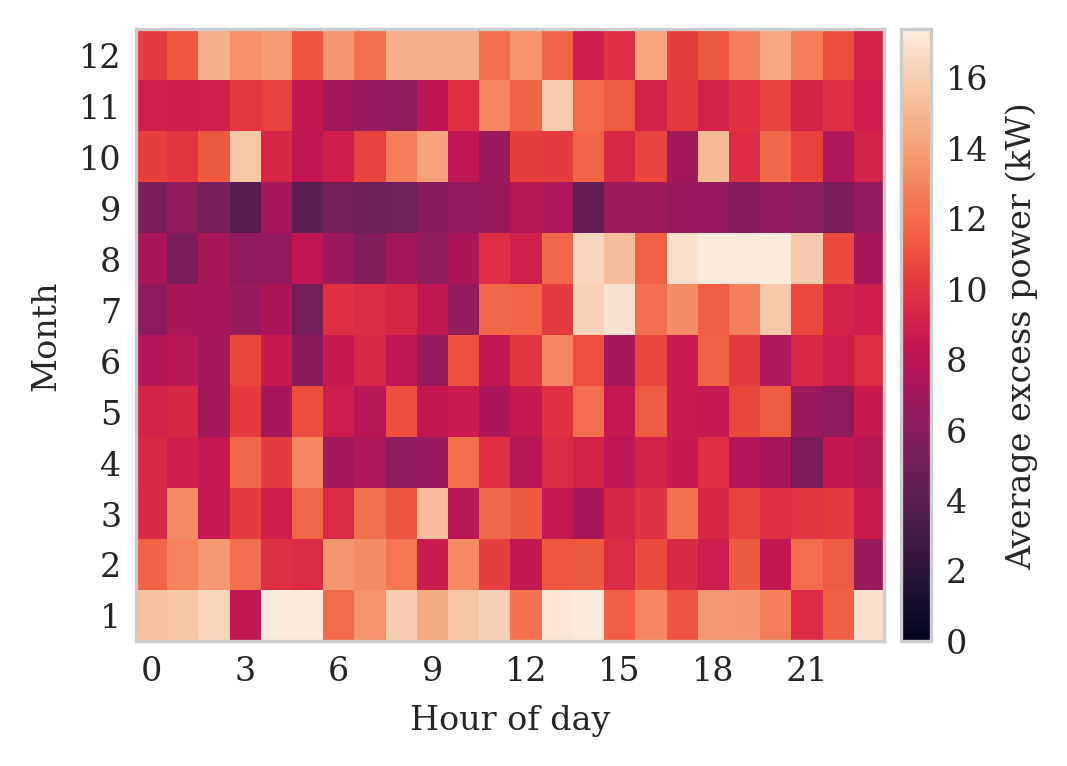}
  \caption{Average excess cooling power per 10-minute interval as a function of month and hour of day.}
  \label{fig:excess_hour_month}
\end{figure}

Seasonal and diurnal structures are summarized in the hour-by-month heatmap in Fig.~\ref{fig:excess_hour_month}. Excess is largest in winter (January) and late-year (December), with a secondary peak in August, while late spring and early fall (May--June, September) show the smallest average excess. In terms of time of day, winter exhibits a pronounced pre-dawn excess band around 05:00, consistent with the interaction between ambient conditions and load-driven cooling demand. These patterns suggest that a sizable share of excess energy is concentrated in specific season-hour combinations rather than being uniformly distributed.

\begin{figure}[h]
  \centering
  \includegraphics[width=0.9\linewidth]{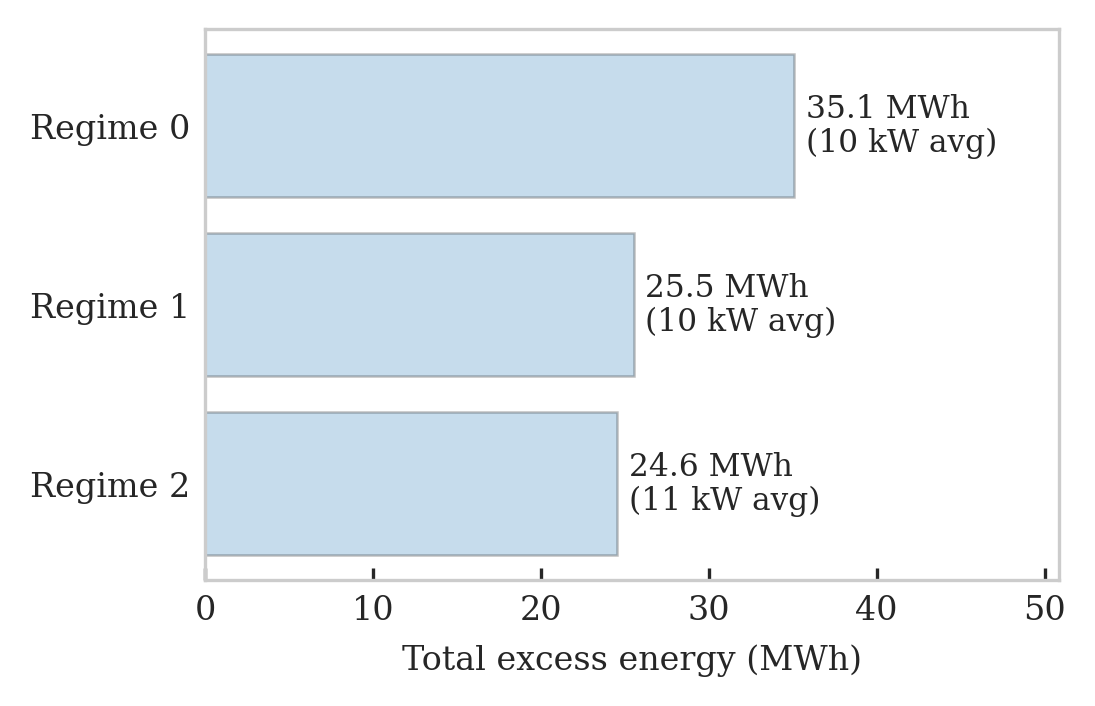}
  \caption{Total excess cooling energy by operating regime, with annotated average excess power per interval.}
  \label{fig:excess_regime}
\end{figure}

Daily roll-ups (not shown) confirm that a small number of ``spiky'' days contribute disproportionately to the annual total, particularly in August. To understand where in the operating space this excess arises, we also group intervals by the three operating regimes introduced in Step~1. Figure~\ref{fig:excess_regime} shows that regime~0 contributes the largest share of total excess energy (about $41\%$), primarily because it occupies the most time, while regime~2 has the highest excess intensity per 10-minute interval (slightly higher mean excess per step than the other regimes). Regime~1 corresponds to the highest average IT power and contributes roughly $30\%$ of the total excess. Median PUE is very similar across regimes (all close to $1.054$), indicating that the excess identified here reflects relatively subtle departures from the surrogate baseline rather than gross efficiency failures.

Overall, Step~2 reveals that (i) the annual excess cooling energy is concentrated in particular months and hours, (ii) a small set of high-excess days accounts for a substantial share of the total, and (iii) certain operating regimes are slightly more waste-intensive than others. These characterizations provide both a quantitative upper bound on recoverable energy and a set of temporal and regime-specific targets for the counterfactual micro-adjustments explored in Step~3.
\subsection{Counterfactual savings and safety}
\label{subsec:step3_savings}

Step~3 uses the Step 1 surrogate as a counterfactual oracle to evaluate small, guardrail-constrained adjustments to $T_{\mathrm{sup}}$ and subloop flows, holding IT load fixed and reusing the same physics-derived features and regime labels. For each 10-minute interval, the algorithm scans a discrete action grid and selects the feasible action that maximizes the one-step reduction in accessory power.

We report three levels of conservatism. With only the physics-guardrails enforced (no cap against Step 2 excess or reviewer filters), the integrated one-step savings over 2023 are $126.8~\mathrm{MWh}$, or about \$7.6\,k at \$60/MWh. When these savings are capped by the Step 2 excess on a per-interval basis, the annual total is $82.1~\mathrm{MWh}$ (96.4\% of the $85.2~\mathrm{MWh}$ excess), corresponding to roughly \$4.9\,k. After applying the full reviewer-oriented diagnostics (in-distribution checks, materiality threshold, and simple hysteresis), 1,791 actions remain (3.6\% of 49,869 intervals), yielding $13.4~\mathrm{MWh}$ (15.8\% of Step 2 excess) and about \$0.81\,k per year. The capped $82.1~\mathrm{MWh}$ thus acts as a practical upper bound, while the reviewer-pass $13.4~\mathrm{MWh}$ is a defensible lower bound on recoverable cooling energy.

\begin{figure*}[h]
  \centering
  \includegraphics[width=0.75\textwidth]{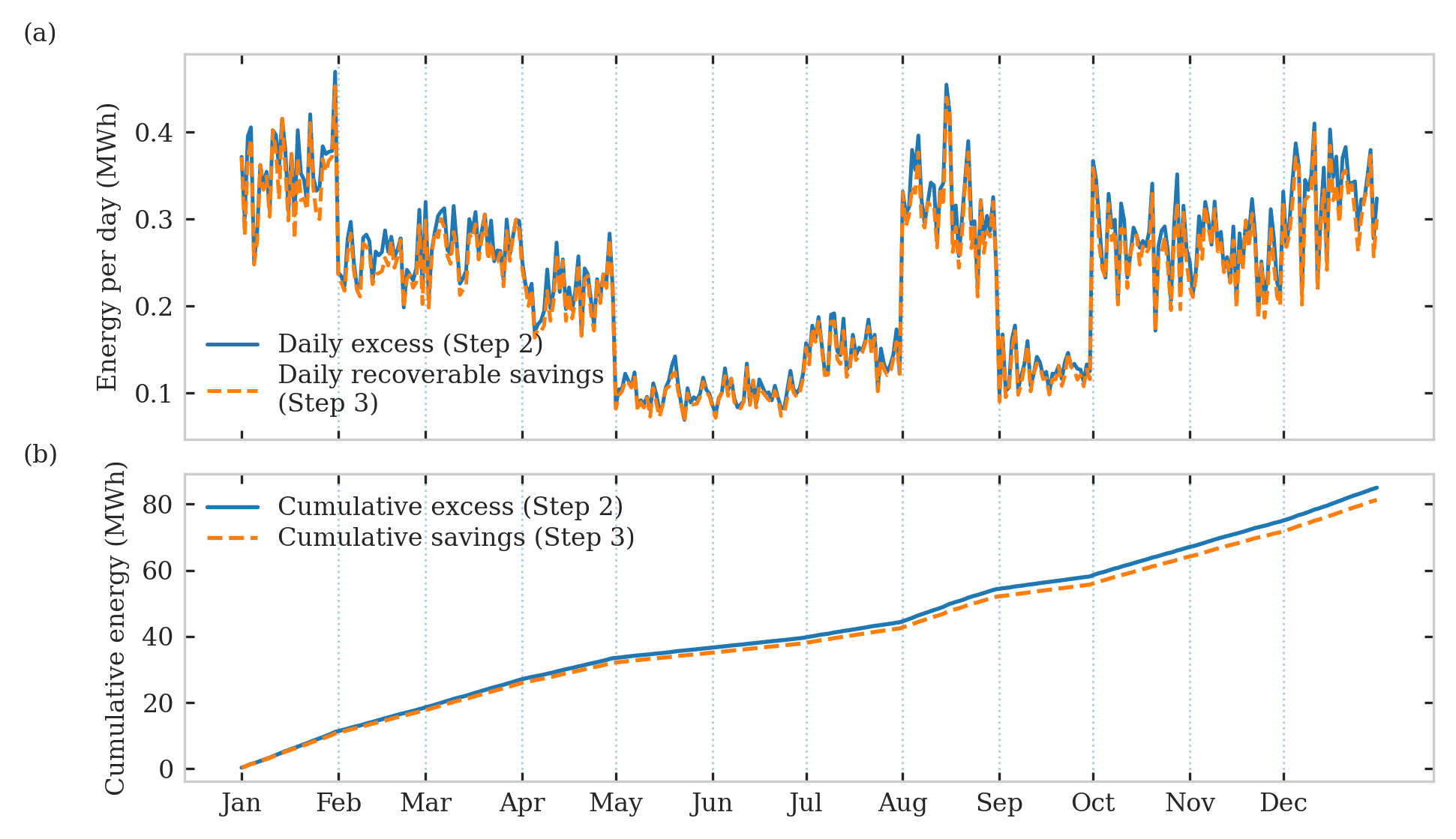}
  \caption{Daily and cumulative accessory-energy profiles over 2023. (a) Daily excess accessory energy (Step~2) and daily recoverable savings (Step~3, capped). (b) Corresponding cumulative sums, showing Step~3 remains below Step~2 while closely tracking it, which illustrates the upper-bound capture potential.}
  \label{fig:fig10}
\end{figure*}

\begin{figure}[h]
  \centering
  \includegraphics[width=0.9\linewidth]{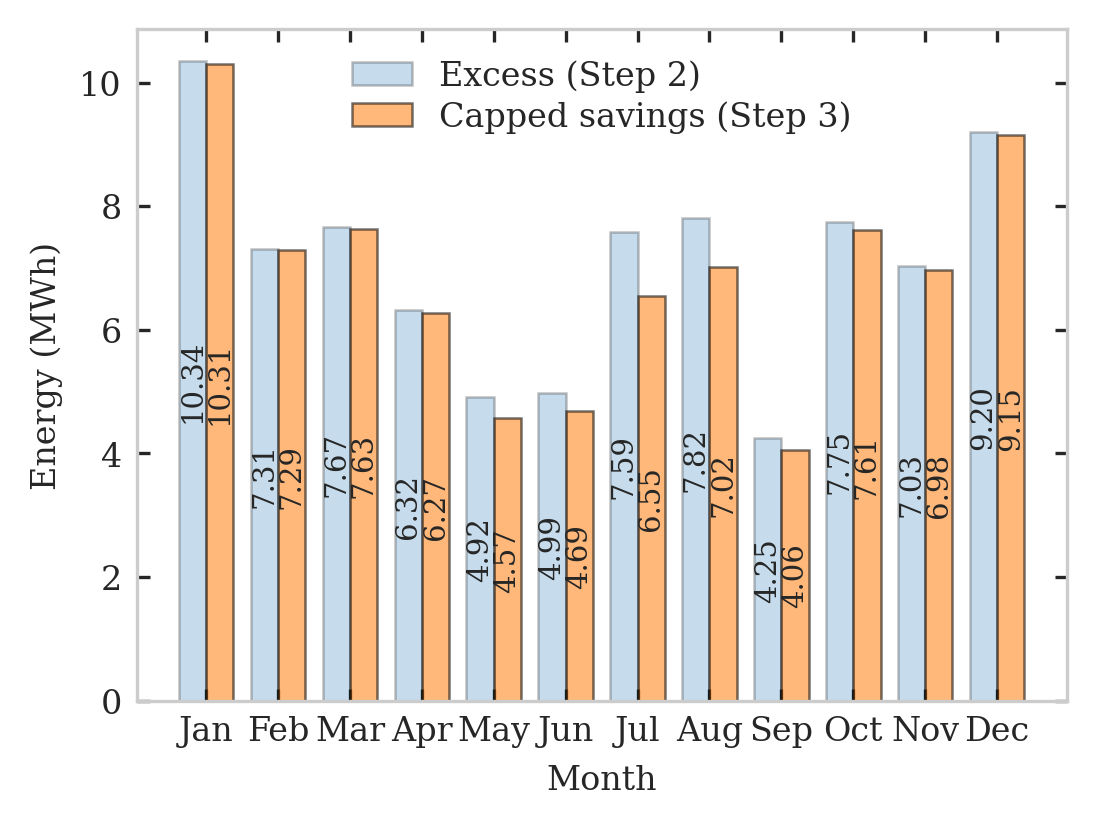}
  \caption{Monthly breakdown of capped Step~3 savings. Savings concentrate in months with large Step~2 excess (January, December, and selected shoulder months), confirming that the counterfactual policy acts on existing temporal hot spots rather than creating artificial savings.}
  \label{fig:fig11}
\end{figure}

Figure~\ref{fig:fig10} summarizes the temporal structure of these savings. For visual clarity, the top panel aggregates the 10-minute results to daily excess and daily recoverable savings. The two curves track each other closely throughout the year: days with high excess in winter and late-year show correspondingly high counterfactual savings, while shoulder periods exhibit smaller values. The lower panel shows the cumulative sums, where the Step 3 curve remains strictly below the Step 2 excess and finishes the year at $\sim 82~\mathrm{MWh}$ versus $\sim 85~\mathrm{MWh}$, illustrating that the policy never ``claims'' more energy than the excess baseline but captures almost all of it in the upper-bound case.

Per-action savings in the reviewer-pass set are modest but nontrivial. The median reduction is $0.035~\mathrm{MW}$ (35~kW), with an interquartile range of $0.026$-$0.052~\mathrm{MW}$, a 95th percentile of $\approx 0.10~\mathrm{MW}$, and a maximum of $\approx 0.26~\mathrm{MW}$. Control moves are similarly small. The mean supply-temperature increase is $0.12^{\circ}\mathrm{C}$ with median $0.0^{\circ}\mathrm{C}$; 53\% of accepted actions keep $T_{\mathrm{sup}}$ unchanged, and among those with $\Delta T_{\mathrm{sup}}>0$ the 95th percentile is $0.4^{\circ}\mathrm{C}$ (maximum $0.6^{\circ}\mathrm{C}$). Dominant-loop flow scales have a mean $0.994$ and a minimum $0.95$, non-dominant loops have a mean $0.977$ and a minimum $0.95$, and no loop is ever reduced below 95\% of baseline.

Figure~\ref{fig:fig11} decomposes the capped savings by month. The largest recoverable energy occurs in January and December, with secondary contributions in March, August, and October, while months with little excess (May-June, September) show correspondingly small savings. This alignment with the Step 2 monthly excess confirms that the counterfactual layer is acting on the same temporal ``hot spots'' identified by the excess analysis rather than creating artificial savings in low-excess periods. Regime- and loop-level summaries (not shown) further indicate that capped savings are split across regimes~0, 1, and~2 (about $34.0$, $22.5$, and $25.6~\mathrm{MWh}$, respectively), with regime~2 exhibiting the highest savings intensity per 10-minute step, and that $\sim 98.8\%$ of capped savings occur when subloop~2 is dominant, consistent with its role as the main heat-rejection path.

To assess the sensitivity of these estimates to surrogate uncertainty, we note that the Step~1 test MAE of 0.026~MW corresponds to an annual energy uncertainty of approximately $\pm 13$~MWh if the error were systematic across all intervals. However, because residuals are approximately symmetric and centered near zero (mean $\approx 0.0007$~MW), random errors largely cancel over aggregation. The capped savings of 82.1~MWh exceed this uncertainty band by a factor of six, providing confidence that the reported savings are not artifacts of model error. For the more conservative reviewer-pass estimate of 13.4~MWh, the margin is smaller, and these savings should be interpreted as order-of-magnitude guidance rather than precise predictions.

All reviewer-pass actions satisfy the explicit guardrails by construction: counterfactual PUE never drops below~1 (minimum $\approx 1.027$), total heat removal remains close to baseline, temperature lifts stay above the imposed minimum, and no flow or temperature bound is violated. In-distribution checks confirm that the counterfactual states lie within the central bulk (1st-99th percentile) of the Step 1 training distribution for key features, and the materiality filter prevents the aggregate $13.4~\mathrm{MWh}$ from being dominated by changes within the surrogate's error band. Overall, Step~3 indicates that safe, interpretable micro-adjustments can realistically recover on the order of $10$-$15~\mathrm{MWh}$ per year under conservative reviewer filters, with an upper limit of $82.1~\mathrm{MWh}$ under the Step 2 excess baseline.

\section{Conclusion}
\label{sec:conclusion}

This work presented a three-stage, physics-guided framework for identifying and quantifying micro-inefficiencies in the liquid-cooling operation of the Frontier exascale system. Using one year of 10-minute Frontier telemetry, we first trained a monotone, regime-aware LightGBM surrogate for accessory power that reproduces $P_{\mathrm{acc}}$ with errors on the order of a few tens of kilowatts and PUE errors of only a few thousandths. This surrogate provides a stable, interpretable baseline that respects basic thermodynamic trends and is accurate enough to support both excess-use detection and counterfactual evaluation. Building on this baseline, Step~2 converts residuals between actual and expected accessory power into a time-resolved estimate of excess cooling energy, revealing about $85~\mathrm{MWh}$ of annual excess and showing that it is concentrated in specific months, hours, and operating regimes rather than being uniformly distributed.

Step~3 then uses the same surrogate and feature space as a counterfactual oracle, exploring small, guardrail-constrained adjustments to supply temperature and subloop flows at each 10-minute interval. Under pure physics guardrails, the counterfactual policy can recover roughly $127~\mathrm{MWh}$ of cooling energy; when capped by the Step 2 excess on a per-interval basis, this shrinks to $\sim 82~\mathrm{MWh}$, or about $96\%$ of the identified excess. After applying conservative reviewer filters (in-distribution checks, materiality thresholds, and simple hysteresis), the resulting sparse policy touches only a few percent of time steps yet still delivers about $13~\mathrm{MWh}$ per year of credible savings through micro-adjustments at the tens-of-kilowatt level. Together, these results demonstrate that safe, interpretable micro-adjustments can recover up to 96\% of identified cooling inefficiencies, corresponding to approximately 82~MWh annually under the Step~2 excess baseline, with a more conservative estimate of 13--15~MWh under strict reviewer filters. These findings suggest that even in a facility with already excellent PUE, there remains a measurable band of recoverable cooling energy accessible through physics-guided setpoint tuning.

The present study is subject to several limitations that motivate future work. First, all results are based on offline counterfactual evaluation of a single year of historical data from one installation; the surrogate and policy are not yet validated under prospectively collected telemetry or in closed-loop operation, and the conclusions would require site-specific recalibration before being applied to other liquid-cooled sites with different hardware, controls, or climates. Second, the action space is deliberately narrow (small $\Delta T_{\mathrm{sup}}$ and modest flow trims), and the guardrails are conservative, so the reported savings should be viewed as lower bounds relative to what more aggressive but still safe policies might achieve. Third, the framework currently treats IT workload as an exogenous input and does not co-optimize scheduling and cooling, nor does it model interactions with upstream plant equipment (e.g., chillers, towers, or heat-recovery systems).

Future work could relax these constraints along several dimensions. On the modeling side, one could explore richer surrogate architectures or multi-output models that jointly predict accessory power, temperatures, and flows while preserving monotonicity and interpretability. On the control side, the micro-actions identified here could serve as building blocks for model-predictive or safe reinforcement-learning controllers that operate online with explicit uncertainty handling and operator-in-the-loop overrides. Finally, extending the analysis to multi-year datasets, additional exascale systems, and scenarios with dynamic tariffs or heat-recovery value streams would clarify how robust the observed savings patterns are and how best to integrate physics-guided micro-tuning into broader HPC sustainability and grid-interaction strategies. 

\section{Code Availability}
The code used in this study can be accessed at \href{https://github.com/m-iml/ML-Optimization-Data-Centers}{https://github.com/m-iml/ML-Optimization-Data-Centers}.

\section*{Acknowledgment}
This research was partially supported by the University of Michigan-Dearborn Office of Research through Research Initiation \& Development (RID) Grant.

\bibliographystyle{IEEEtran}
\bibliography{ref}

\end{document}